\documentclass[3p]{elsarticle}
\usepackage{natbib}
\usepackage{lineno,hyperref}
\usepackage{latexsym,amsmath,amssymb,amsthm,amsfonts,graphicx,graphics,epsfig,url,color}
\usepackage{float}
\usepackage{mcode}
\usepackage{pdflscape}
\usepackage{afterpage}
\modulolinenumbers[5]
\newcommand{\mbf}{\mathbf}

\journal{Physica D}
\makeatletter
\def\ps@pprintTitle{%
 \let\@oddhead\@empty
 \let\@evenhead\@empty
 \def\@oddfoot{\centerline{\thepage}}%
 \let\@evenfoot\@oddfoot}
\makeatother

\begin{document}

\begin{frontmatter}

\title{Phase Distribution Control of a Population of Oscillators}

\author[mymainaddress]{Bharat Monga\corref{mycorrespondingauthor}}
\ead{monga@ucsb.edu}

\author[mymainaddress,mysecondaryaddress]{Jeff Moehlis}
\cortext[mycorrespondingauthor]{Corresponding author}
\ead{moehlis@engineering.ucsb.edu}

\address[mymainaddress]{Department of Mechanical Engineering, Engineering II Building, University of California Santa Barbara, Santa Barbara, CA 93106}
\address[mysecondaryaddress]{Program in Dynamical Neuroscience, University of California Santa Barbara, Santa Barbara, CA 93106, United States}

\begin{abstract}
The collective behavior of biological oscillators has been recognized as an important problem for several decades, but its control has come into limelight only recently. Much of the focus for control has been on desynchronization of an oscillator population, motivated by the pathological neural synchrony present in essential and parkinsonian tremor. Other applications, such as the beating of the heart and insulin secretion, require synchronization, and recently there has been interest in forming clusters within an oscillator population as well. In this article, we use a formulation that allows us to devise control frameworks to achieve all of these distinct collective behaviors observed in biological oscillators. This is based on the Fourier decomposition of the partial differential equation governing the evolution of the phase distribution of a population of identical, uncoupled oscillators. Our first two control algorithms are Lyapunov-based, which work by decreasing a positive definite Lyapunov function towards zero. Our third control is an optimal control algorithm, which minimizes the control energy consumption while achieving the desired collective behavior of an oscillator population. Motivated by pathological neural synchrony, we apply our control to desynchronize an initially synchronized neural population. Given the proposed importance of enhancing spike time dependent plasticity to stabilize neural clusters and counteract pathological neural synchronization, we formulate the phase difference distribution in terms of the phase distribution, and prove some of its fundamental properties, and in turn apply our control to transform the neural phase distribution to form clusters. Finally, motivated by eliminating cardiac alternans, we apply our control to phase shift a synchronous cardiac pacemaker cell population. For the systems considered in this paper, the control algorithms can be applied to achieve any desired traveling-wave phase distribution, as long as the combination of the initial phase distribution and phase response curve is non-degenerate.  To demonstrate the effectiveness of our control for each of these applications, we show that a population of 100 phase oscillators with the applied control mimics the desired phase distribution.

\end{abstract}


\end{frontmatter}


\section{Introduction}
Populations of nonlinear oscillators are found in a variety of applications from physics, chemistry, biology, and engineering \cite{physics,kuramoto,winfree,Keener2009}. The collective behavior of such oscillators varies, and includes synchronization, desynchronization, and clustering. For example, synchronization in beta cells is crucial for efficient insulin secretion \cite{beta_synch}, the beating of the heart is regulated by constant pacing of synchronized cardiac pacemaker cells \cite{sn_synch1,sn_synch2}, and neural synchrony is essential in visual and odor processing \cite{visual,odor}, and also in learning and memory recall \cite{learn,memory}. However, synchronization can be detrimental as well. For example, pathological neural synchronization in the thalamus and the subthalamic nucleus (STN) brain region is hypothesized to be one of the causes of motor symptoms for essential and parkinsonian tremor, respectively~\cite{thalamus_syncro,STN_syncro}; this motivates the goal of designing a control input to desynchronize an oscillator population. Recently there has also been focus on achieving partial synchrony through clustering instead of complete neural desynchronization \cite{tim,Matchen2018,clustered}. One motivation behind such clustering is to rewire neural connections by enhancing spike time dependent plasticity which potentiates intra-cluster synaptic connections and depresses inter-cluster connections. This potentially helps in long-term stabilizability of the clusters in the presence of noise.

Such diversity of collective behavior has motivated researchers to develop specific control techniques to achieve different behaviors. For example, \cite{circ1,circ2,Kuritz2018} develop control to promote synchrony, \cite{tass2003,wilson11,dan14} develop control to promote desynchronization, and \cite{tim,Matchen2018} develop ways to promote clustering. We note that some of these previously proposed algorithms to promote collective behavior are based on individual neuron models \cite{moehlis06,danzl,nabi2010,dan14}, and some can face implementation challenges if they require observability of phases of all neurons at all times \cite{nabi2010}, or demand initial phases to be sufficiently close \cite{dan14,tass2007}.  There are also population-level algorithms for desynchronization in the literature which use multiple inputs \cite{tass2001,tass2003,tass2007}, making experimental implementation challenging because they require multiple electrodes to be implanted in a small region of brain tissue.  

In this article we overcome these difficulties by developing unified control frameworks which can achieve all of the collective behaviors mentioned above using a single control input. Our algorithms are based on phase reduction, a classical reduction technique based on isochrons~\cite{guckenheimer1975}, which has been instrumental in the development of many of the above control algorithms. It reduces the dimensionality of a dynamical system with a periodic orbit to a single phase variable, and captures the oscillator's phase change due to an external perturbation through the phase response curve (PRC). This can make the analysis of high dimensional systems more tractable, and their control \cite{moehlis06,dan14, zlotnik, tass2007,prc_exp2,monga} experimentally implementable; see e.g., \cite{stigen,nabi2013,snari15,zlotnik}.

The algorithms presented in this paper use a partial differential equation (PDE) formulation which governs the evolution of the probability distribution of phases (phase distribution) of a population of identical, uncoupled oscillators \cite{brow04,tass2007}. We use Fourier analysis to decompose this PDE into a system of ordinary differential equations (ODEs) governing the evolution of the Fourier coefficients of the phase distribution. Thus, to transform the phase distribution of an oscillator population to a desired distribution, we drive the corresponding Fourier coefficients to the Fourier coefficients of the desired distribution. Our first two algorithms are Lyapunov-based, which work by decreasing the $L^2$ norm difference between the current and the desired phase distributions. Note that a related control algorithm has been published in \cite{monga18}, where we did not employ Fourier analysis to decompose the PDE into a systems of ODEs, but like the present first two algorithms it also decreases the $L^2$ norm difference between the current and the desired phase distributions. This formulation in Fourier space makes our algorithm suitable for using a pseudospectral method for more accurate numerical simulation of the PDEs, which enables us to realize new control objectives and applications discussed in Section \ref{app}. Such a formulation also allows us to obtain the degenerate set of phase distributions and phase response curves for which the control would not work. For the control formulation in \cite{monga18}, we employed a method of lines type approach for numerically simulating our PDEs, but numerical dissipation present in this approach limited the versatility of control, especially when going from a uniform phase distribution to a synchronous distribution.

Our third algorithm is an optimal control algorithm, which unlike the previous two algorithms, minimizes the control energy consumption while achieving the desired control objective. We formulate it by constructing a cost function in terms of the Fourier coefficients of the phase distribution. This formulation results in high dimensional Euler-Lagrange equations that we solve as a two point Boundary Value Problem (BVP) numerically. Since the BVP is high dimensional, we construct a modified Newton Iteration method that is effective for our problem. To demonstrate the effectiveness of our control algorithms for each of the applications considered, we apply them to a population of 100 uncoupled phase oscillators, and show that the population of phase oscillators with the applied control mimics the desired phase distribution. Other control algorithms based on the probability distribution of phases include \cite{dan14,dan_pde,8467360}.

This article in organized as follows. In Section \ref{back}, we give background on phase reduction, and the partial differential equation for the phase distribution. In Section \ref{con_alg}, we develop a pseudospectral framework to write distributions as a finite Fourier series, and devise a Lyapunov-based control algorithm to control their Fourier coefficients. We construct a degenerate set of phase distributions and phase response curves for which the devised control would not work in Section \ref{degeneracy}. In Section \ref{numerics}, we detail the pseudospectral method used for numerical simulations throughout the article. In Sections \ref{app}, we demonstrate versatility of our control through several diverse applications and show the corresponding simulation results. In the same section, we formulate the phase difference distribution and prove some of its properties. In Section \ref{coup_noise}, we devise another Lyapunov-based control to take into account the effect of white noise on the oscillator population. We develop our optimal control algorithm in Section \ref{optimal_control} and compare it with our Lyapunov-based algorithm. Section \ref{con} summarizes our work and concludes by suggesting future extensions and tools needed for experimental implementation of our algorithms. \ref{model_param} lists the mathematical models used in this article. The modified Newton Iteration method for solving a high dimensional BVP is detailed in \ref{bvp_mni}. 

We note that beyond using a formulation in terms of Fourier series, other improvements and extensions with respect to \cite{monga18} include the formulation of the degenerate set, the use of a pseudospectal method for more accurate numerical simulations, formulation of the phase difference distribution, novel applications including the incorporation of the phase difference distribution and plasticity into the control set-up, extension of the control algorithm to account for the presence of noise, and the formulation of an optimal control algorithm.

\section{Background}\label{back}
In this section, we give background on the key concepts of phase reduction, phase response curves, and the partial differential equation for the evolution of the phase density. These will be crucial for the formulation of our control algorithms in Section \ref{con_alg}.

\subsection{Phase Reduction}
Phase reduction is a classical technique to describe the dynamics near a periodic orbit. It works by reducing the dimensionality of a dynamical system to a single phase variable $\theta$~\cite{winfree,kuramoto}. Consider a general $n$-dimensional dynamical system given by 
\begin{equation}
\frac{d \mbf{x}}{dt} = F(\mbf{x}), \qquad \mbf{x} \in \mathbb{R}^n, \qquad (n \ge 2) .
\label{dxdt}
\end{equation}
Suppose this system has a stable periodic orbit $\gamma(t)$ with period $T$. For each point $\mbf{x}^*$ in the basin of attraction of the periodic orbit, there exists a corresponding phase $\theta(\mbf{x}^*)$ such that
\begin{equation}
\lim_{t \rightarrow \infty} \left| \mbf{x}(t) - \gamma \left( t+\frac{T}{2 \pi} \; \theta(\mbf{x}^*) \right) \right| = 0,
\end{equation}
where $\mbf{x}(t)$ is the flow of the initial point $\mbf{x}^*$ under the given vector field. The function $\theta(\mbf{x})$ is called the {\it asymptotic phase} of $\mbf{x}$, and takes values in 
$[0, 2 \pi)$.  For neuroscience applications, we typically take $\theta=0$ to correspond to the neuron firing an action potential.  {\it Isochrons} are level sets of this phase function, and it is typical to define isochrons so that the phase of a trajectory advances linearly in time both on and off the periodic orbit, which implies that
\begin{equation}
\frac{d\theta}{dt} = \frac{2 \pi}{T} \equiv \omega
\label{theta_flow}
\end{equation}
in the entire basin of attraction of the periodic orbit. Now consider the system
\begin{equation}
\frac{d \mbf{x}}{dt} = F(\mbf{x})+U(t), \qquad \mbf{x} \in \mathbb{R}^n,
\label{udxdt}
\end{equation}
where $U(t) \in \mathbb{R}^n$ is an infinitesimal control input. Phase reduction can be used to reduce this system to a one-dimensional system given by \cite{winf01,kura84,brow04,tutorial}:
\begin{equation}
\dot \theta = \omega + U(t)^T\mathcal{Z}(\theta).
\end{equation}
Here $\mathcal{Z}(\theta) \equiv \nabla_{\gamma(t)} \theta \in \mathbb{R}^n$ is the gradient of phase variable $\theta$ evaluated on the periodic orbit and is referred to as the {\it (infinitesimal) phase response curve (PRC)}. It quantifies the effect of an infinitesimal control input on the phase of a periodic orbit. 

In this article we consider control inputs of the form $U(t)= [u(t), \ 0,\ldots ,0]^T$. This comes into phase reduction as $\dot \theta = \omega +\mathcal{Z}_1(\theta)u(t)$, where $\mathcal{Z}_1(\theta)$ is the first  component of the PRC.  Without loss of generality, we will do away with the subscripts and write the first component of PRC as $\mathcal{Z}(\theta)$. Thus the phase reduction is written as
\begin{equation}
\dot \theta = \omega + \mathcal{Z}(\theta)u(t).
\label{spr}
\end{equation}
Note that such a control input is motivated by the applications we consider in this article, where only one of the elements of the state vector is affected directly by the control input. The control algorithm in this article can be formulated for a more general control input as well, but as a matter of convenience, we only consider control input of the above form.
\subsection{Phase density equation}
Given a population of noise-free, identical, uncoupled oscillators all receiving the same control input, it is convenient to represent the population dynamics in terms of its probability distribution $\rho(\theta,t)$, with the interpretation that $\rho(\theta,t) d\theta$ is the probability that an oscillator's phase lies in the interval $[\theta,\theta + d \theta)$ at time $t$.  This evolves according to the advection equation \cite{brow04,tass2007,monga18}
\begin{equation}
\frac{\partial \rho(\theta,t)}{\partial t}=-\frac{\partial }{\partial \theta}\left[\left(\omega+\mathcal{Z}(\theta)u(t)\right)\rho(\theta,t)\right].\label{fpe}
\end{equation}
The desired final probability distribution $\rho_f(\theta,t)$ will be taken to be a traveling wave which evolves according to \cite{monga18}
\begin{equation}
\frac{\partial \rho_f(\theta,t)}{\partial t}=-\omega\frac{\partial \rho_f(\theta,t)}{\partial \theta}.\label{dfpe}
\end{equation}
Note that (\ref{dfpe}) is of the same form as (\ref{fpe}) with $u(t)=0$. Since these are probability distributions, it is necessary that $\int_0^{2\pi}\rho(\theta,t)d\theta=\int_0^{2\pi}\rho_f(\theta,t)d\theta=1$

In Section \ref{con_alg}, we will show how these two equations can be used to devise our control algorithms.

\section{Control Algorithm}\label{con_alg}
In this section, we devise a control algorithm to change the probability distribution of a population of oscillators. We do this by approximating the probability distribution as a finite Fourier series and controlling its Fourier coefficients. This algorithm can be applied to a network of noise-free, identical, uncoupled oscillators to achieve any desired traveling-wave probability distribution, as long as the combination of phase distributions and the phase response curve is non-degenerate. A related control algorithm was given in \cite{monga18}, but here we formulate the algorithm in terms of Fourier coefficients; this is better suited for determining the control input using a pseudospectral method which does not have numerical dissipation unlike the method of lines approach used in \cite{monga18}. 

\subsection{Fourier Decomposition}
To devise our control laws, we use the approximation of a finite Fourier series to write the phase distributions and the PRC as
\begin{eqnarray}
\rho(\theta,t)&=&\frac{1}{2\pi}+\sum_{l=1}^{N-1} \left[A_l(t)\cos (l\theta) +B_l(t) \sin (l \theta)\right],\label{rho}\\
\rho_f(\theta,t)&=&\frac{1}{2\pi}+\sum_{l=1}^{N-1}\left[ \widetilde{A_l}(t)\cos (l\theta) + \widetilde{B_l}(t) \sin (l \theta)\right] ,\label{rhof}\\
\mathcal{Z}(\theta)&=&C_0+\sum_{l=1}^{N-1} \left[C_l\cos (l\theta) +S_l \sin (l \theta)\right] .\label{prc}
\end{eqnarray}
Here $N$ is a large number, so the effect of the omitted higher order Fourier modes is negligible. Writing the distribution this way automatically ensures that the phase distribution is $2\pi$-periodic, and that the total probability $\int_0^{2\pi}\rho(\theta,t)d\theta=1$ at all times. Multiplying equation (\ref{rho}) by $\cos (k \theta)$ and $\sin (k\theta)$ on both sides and integrating from $0$ to $2\pi$ with respect to $\theta$, we obtain
\begin{eqnarray*}
A_k(t)&=&\frac{1}{\pi}\int_0^{2\pi}\rho(\theta,t)\cos (k\theta) d\theta,\\
 B_k(t)&=&\frac{1}{\pi}\int_0^{2\pi}\rho(\theta,t)\sin (k\theta) d\theta.
\end{eqnarray*}
Taking the derivative with respect to time of the above equations,
\begin{eqnarray*}
\dot A_k(t)&=&\frac{1}{\pi}\int_0^{2\pi}\frac{\partial \rho(\theta,t)}{\partial t}\cos (k\theta) d\theta=-\frac{1}{\pi}\int_0^{2\pi}\frac{\partial}{\partial \theta} \left[\left(\omega+\mathcal{Z}(\theta)u(t)\right)\rho(\theta,t)\right]\cos (k\theta) d\theta,\\
 \dot B_k(t)&=&\frac{1}{\pi}\int_0^{2\pi}\frac{\partial \rho(\theta,t)}{\partial t}\sin (k\theta) d\theta=-\frac{1}{\pi}\int_0^{2\pi}\frac{\partial}{\partial \theta} \left[\left(\omega+\mathcal{Z}(\theta)u(t)\right)\rho(\theta,t)\right]\sin (k\theta) d\theta.
\end{eqnarray*}
Integrating these equations by parts and imposing periodic boundary conditions, we obtain
\begin{eqnarray}
\dot A_k(t)&=&-k\omega B_k - \mathcal{I}_{kA}(t)u(t),\label{ak}\\
\dot B_k(t)&=&k\omega A_k +\mathcal{I}_{kB}(t)u(t),\label{bk}
\end{eqnarray}
where 
\begin{eqnarray}
\mathcal{I}_{kA}(t)&=&\frac{k}{\pi}\int_0^{2\pi} \mathcal{Z}(\theta)\rho(\theta,t)\sin (k\theta) d\theta,\label{ika}\\
\mathcal{I}_{kB}(t)&=&\frac{k}{\pi}\int_0^{2\pi} \mathcal{Z}(\theta)\rho(\theta,t)\cos (k\theta) d\theta.\label{ikb}
\end{eqnarray}
Similarly we obtain following equations for time derivatives of $\widetilde{A_k}$ and $\widetilde{B_k}$:
\begin{eqnarray}
\dot {\widetilde{A_k}}(t)&=&-k\omega \widetilde{B_k}(t) ,\label{afk}\\
\dot {\widetilde{B_k}}(t)&=&k\omega \widetilde{A_k} (t).\label{bfk}
\end{eqnarray}

\subsection{Control Design}\label{control_design}
If for all $k$, $A_k(\tau)=\widetilde{A_k}(\tau)$ and $B_k(\tau)=\widetilde{B_k}(\tau)$, the phase distribution $\rho$ would be equal to the desired distribution $\rho_f$ at time $\tau$. This motivates us to take our Lyapunov function as the sum of the squared differences of the Fourier coefficients of the current and the desired distribution:
\begin{equation}
V(t)=\frac{1}{2}\sum_{k=1}^{N-1}\left[\left(A_k(t)-\widetilde{A_k}(t)\right)^2+\left(B_k(t)-\widetilde{B_k}(t)\right)^2\right].\label{lf}
\end{equation}
Thus the Lyapunov function is non-negative, and is zero only when $\rho(\theta,t)=\rho_f(\theta,t)$. Its derivative in time evolves as
\begin{equation}
\dot V(t)=I(t)u(t),\label{dotofv}
\end{equation}
where $I(t)$ is given by the sum
\begin{equation}
I(t)=\sum_{k=1}^{N-1}\left[\left(B_k(t)-\widetilde{B_k}(t)\right)\mathcal{I}_{kB}(t) - \left(A_k(t)-\widetilde{A_k}(t)\right)\mathcal{I}_{kA}(t)\right].\label{integrall}
\end{equation}
Then by taking the control input $u(t)=-PI(t)$, where $P$ is a positive scalar, we get the time derivative of the Lyapunov function, $\dot V(t)=-PI(t)^2$ as a negative scalar. Thus, according to the Lyapunov theorem, the control law $u(t)=-PI(t)$ will decrease the Lyapunov function until the current probability distribution becomes equal to the desired distribution. Here we do not consider the degenerate systems where $I(t)\equiv 0$ when $\rho(\theta,t)\ne \rho_f(\theta,t)$ (see Section \ref{degeneracy} for such a system). 

For both experimental and numerical reasons, it is more practical to have a bounded control input, so we take a ``clipped'' proportional control law
\begin{eqnarray}
u(t)= \max \left( \min \left(u_{max},-PI(t)\right), u_{min}\right).  \label{pc}
\end{eqnarray}
Here $u_{max}$ and $u_{min}$ are the upper and lower bounds of the control input, respectively. The $\max$, and $\min$ operators find the maximum and minimum of two scalars, respectively.

\section{Degenerate Set}\label{degeneracy}
Note that for certain systems where $\rho(\theta,t)\ne \rho_f(\theta,t)$, equation (\ref{integrall}) gives $I(t)\equiv 0$ for all time $t$, and the probability distribution $\rho(\theta,t)$ would not converge to the desired distribution $\rho_f(\theta,t)$. Here we derive the set of such phase distributions and PRCs that leads to this degeneracy, and give an example of such a system. 

We can re-write $I(t)$ as
\begin{eqnarray}
I(t)&=&\frac{1}{\pi}\int_0^{2\pi}\sum_{k=1}^{N-1}k\left[\left(B_k(t)-\widetilde{B_k}(t)\right)\cos (k\theta) -\left(A_k(t)-\widetilde{A_k}(t)\right)\sin(k\theta)\right] \mathcal{Z}(\theta)\rho(\theta,t)d\theta\nonumber\\
&=&\frac{1}{\pi}\int_0^{2\pi}\left( \frac{\partial \rho}{\partial \theta} - \frac{\partial \rho_f}{\partial \theta} \right) \mathcal{Z}(\theta)\rho(\theta,t)d\theta.\label{dg1}
\end{eqnarray}
Now expanding $\rho(\theta,t), \rho_f(\theta,t)$, and $\mathcal{Z}(\theta)$ into their complex Fourier series,
\begin{eqnarray}
\rho(\theta,t)&=&\sum_{k=1-N}^{N-1} a_k(t)\exp(ik\theta), \quad \rho_f(\theta,t)=\sum_{k=1-N}^{N-1} \widetilde{a_k}(t)\exp(ik\theta),\nonumber\\
\mathcal{Z}(\theta)&=&\sum_{k=1-N}^{N-1} c_k\exp(ik\theta),\nonumber
\end{eqnarray}
where
\begin{eqnarray}
 a_{\pm k}(t)&=&\frac{A_k(t)\mp i B_k(t)}{2},\quad  \widetilde{a_{\pm k}}(t)=\frac{\widetilde{A_k}(t)\mp i \widetilde{B_k}(t)}{2},\nonumber \\ c_{\pm k}&=&\frac{C_k\mp i S_k}{2}, \quad k=1,\ldots,N-1,\nonumber\\  a_0(t)&=&A_0(t), \quad  \widetilde{a_0}(t)=\widetilde{A_0}(t), \quad c_0(t)=C_0(t), \nonumber
\end{eqnarray}
we can write $I(t)$ from equation (\ref{dg1}) as
\begin{eqnarray}
I(t)=\sum_{p=1-N}^{N-1}\sum_{q=1-N}^{N-1}\sum_{r=1-N}^{N-1}\left[ip\left(a_p(t)-\widetilde{a_p}(t)\right) c_q a_r(t) \frac{1}{\pi}\int_0^{2\pi}\exp\left(i(p+q+r)\theta\right)d \theta\right].\label{dg3}
\end{eqnarray}
Thus the degenerate set of phase distributions and PRCs can be written in terms of their respective Fourier coefficients as
\begin{eqnarray}
\sum_{p \in M}\sum_{q=1-N}^{N-1}\sum_{r=1-N}^{N-1}\left[i2p\left(a_p(t)-\widetilde{a_p}(t)\right) c_q a_r(t)\delta_{p+q+r,0}\right]\equiv 0\label{dg2}
\end{eqnarray}
for all time $t$, where $M$ is the subset of integers ranging from $1-N$ to $N-1$ for which $a_p(t)\ne \widetilde{a_p}(t)$, and $\delta_{p+q+r,0}$ is the Kronecker delta, which is equal to $1 \ \forall \ p+q+r=0$, and is $0$ otherwise.

\subsection{Degenerate System Example}
As an example degenerate system, we consider the Type I PRC near a SNIPER bifurcation given by \cite{brow04}
\begin{equation}
\mathcal{Z}(\theta)=\frac{2}{\omega}\left(1-\cos(\theta)\right) \nonumber.
\end{equation}
Thus $c_0=2/\omega, \ c_{\pm1}=1/\omega$, while rest of the PRC Fourier coefficients are $0$. We take the desired distribution as a uniform distribution,
\begin{equation}
\rho_f(\theta,t)=\frac{1}{2\pi}.\nonumber
\end{equation}
Thus $\widetilde{a_0}(t)= 1/2\pi$, while rest of the Fourier coefficients for $\rho_f$ are $0$ for all times. For the degenerate set, $I \equiv 0$, and thus $\rho(\theta,t)$ is a traveling wave moving in the $+\theta$ direction. We take it as
\begin{equation}
\rho(\theta,t)=\frac{\sin^2(\theta-\omega t)}{\pi}\nonumber.
\end{equation}
It is a physically realistic distribution since $\rho(\theta,t) \ge 0$, and $\int_0^{2\pi}\rho(\theta,t)d\theta=1$. Thus $a_0(t)=1/2\pi, \ a_{\pm2}(t)=-\exp(\mp i2\omega t)/4\pi$, while rest of its Fourier coefficients are $0$. 

There are only two nonzero cases to consider in the summation of the degenerate set (equation (\ref{dg2})):
\begin{eqnarray*}
p=2, q=0, r=-2; \qquad i2(2)\left(-\frac{\exp(-i2\omega t)}{4\pi}-0\right)\left(\frac{2}{\omega}\right)\left(-\frac{\exp(i2\omega t)}{4\pi}\right)=\frac{i}{2\omega\pi^2},\\
p=-2, q=0, r=2; \qquad -i2(2)\left(-\frac{\exp(i2\omega t)}{4\pi}-0\right)\left(\frac{2}{\omega}\right)\left(-\frac{\exp(-i2\omega t)}{4\pi}\right)=-\frac{i}{2\omega\pi^2},\\
I(t)=\frac{i}{2\omega\pi^2}-\frac{i}{2\omega\pi^2}=0.
\end{eqnarray*}
Thus as $I$ is zero even though the phase distributions are not equal, this is a degenerate system. This can also be verified by analytically evaluating the integral in equation (\ref{dg1}) to be zero, i.e.,
\begin{eqnarray*}
I(t)&=&\frac{4}{\omega\pi^3}\int_0^{2\pi}\sin^3(\theta-\omega t) \cos (\theta -\omega t)(1-\cos \theta)d\theta\\
&=&\left.\frac{4}{\omega\pi^3}\left[\frac{\cos(\theta - 2\omega t )-\cos(2\theta-2\omega t )}{8} + \frac{\cos( 3\theta-2\omega t )}{24} - \frac{\cos(3\theta-4\omega t )}{48}\right.\right.\\
 &&\qquad \qquad \qquad \qquad \qquad \qquad \left.\left. + \frac{ \cos(4\theta-4\omega t )}{32} -\frac{ \cos(5\theta-4\omega t)}{80} + \frac{3}{32}\right]\right|_0^{2\pi}=0.
\end{eqnarray*}
Note that such degeneracy arises due to the inherent simplicity and symmetry present in the system under consideration, and thus should not be considered a limitation of the devised control law. ``Real world'' systems would have more than two Fourier modes and some sort of asymmetry, which would avoid degeneracy.

\section{Numerical Methods}\label{numerics}
Here we give details on the numerical methods we use for the simulation results that we present in the next section. Since we are dealing with periodic probability distributions in $\theta$, we take the initial (and, later, the desired distributions) as a von Mises distribution \cite{von_dis}
\begin{eqnarray}
\rho(\theta,0)=\frac{e^{\kappa \cos (\theta+\theta_0)}}{2\pi \mathcal{I}_0(\kappa)},\label{von}
\end{eqnarray}
with $\mathcal{I}_0(\kappa)$ the modified Bessel function of first kind of order 0. For such a distribution, the mean is $\theta_0$, and the variance is $1-\mathcal{I}_1(\kappa)/\mathcal{I}_0(\kappa)$, where $\mathcal{I}_1(\kappa)$ is the modified Bessel function of first kind of order 1. The variance decreases as $\kappa$ increases, and so the distribution becomes narrower and taller. To demonstrate the effectiveness of our control algorithm, we apply the control input given by equation (\ref{pc}) to a population of 100 phase oscillators
\begin{equation}
\dot \Theta_i(t)=\omega+\mathcal{Z}(\Theta_i(t))u(t),\quad i=1,2,\ldots,100.
\end{equation}
For the case where initial distribution $\rho(\theta,0)$ is a uniform distribution ($\kappa=0$), we take the initial value of phase oscillators $\Theta_i(0)=2\pi(i-1)/100$, and for a non-zero $\kappa$, we use the command \mcode{randraw('vonmises', [Theta_0, kappa], 100 )} from the  circular statistical toolbox developed for Matlab in \cite{toolbox} to initialize the phase oscillators corresponding to a von Mises distribution (equation (\ref{von})).

We discretize $\theta$ into a uniform mesh with $2N=128$ grid points. We choose this grid size for a good spatial resolution of the probability distribution, and efficient computation of the fast Fourier transform algorithm. For computing the PRCs of the various models presented in next section, we use the XPP package \cite{xpp} with a time step $T/N$.  We scale the PRC computed by this package by $\omega$, as we consider PRC as $\mathcal{Z}(\theta)=\frac{\partial \theta}{\partial \mbf{x}}$, whereas the computed PRC from the XPP package is $\tilde{\mathcal{Z}}(t)=\frac{\partial t}{\partial \mbf{x}}$ \cite{brow04,tutorial}. Then we use the Matlab command \mcode{fft} to compute the Fourier coefficients of the initial distribution $\rho(\theta,0)$. Note that the \mcode{fft} command computes coefficients of the complex Fourier series given as
\begin{equation*}
\rho(\theta,0)=\sum_{k=1-N}^{N} a_k(0)\exp(ik\theta),
\end{equation*}
giving an output $[a_0, a_1(0),\ldots, a_N(0),a_{1-N}(0),a_{2-N}(0),\ldots,a_{-1}(0)]\times 2N$. From these coefficients, we then compute the coefficients of the real Fourier series
\begin{equation*}
\rho(\theta,0)=A_0+\sum_{k=1}^{N-1} A_k(0)\cos (k\theta) +B_k(0) \sin (k \theta),
\end{equation*}
as 
\begin{eqnarray*}
A_0&=&a_0,\\
A_k(0)&=&(a_k(0)+a_{-k}(0)),\\
B_k(0)&=&i(a_k(0)-a_{-k}(0)).
\end{eqnarray*}
The same procedure is adopted to compute real Fourier coefficients of $\rho_f(\theta,0)$ ($\widetilde{A_k}(0), \ \widetilde{B_k}(0))$ and the PRC ($C_k, \ S_k)$.

To evolve these coefficients over time, ODEs given by equations (\ref{ak}), (\ref{bk}), (\ref{afk}), (\ref{bfk}) are evolved in time using a fourth order Runge-Kutta method with a fixed time step $dt=T/(8N)$. In order to maintain spectral accuracy, the integrals given by equations (\ref{ika})-(\ref{ikb}) are evaluated in Fourier space, i.e., we take the FFT of the integrand using Matlab command \mcode{fft}, and divide the first term of FFT by $N$ to get the numerical value of the integral at every time step.

\section{Applications}\label{app}
In this section, we apply the control law devised in the previous section to manipulate the population density of uncoupled noise-free oscillators to achieve control objectives in a diversity of applications. These applications are desynchronizing an initially synchronized neuron population for the treatment of parkinsonian and essential tremor, forming neuron clusters from an initial desynchronized neuron population to maximize neural plasticity, and eliminating cardiac alternans by phase shifting a synchronized population of cardiac pacemaker cells. For all these applications, we consider underactuated dynamical systems with only one degree of actuation: the control input vector is $U(t) = [u(t), \ 0,\ldots ,0]^T$. We make this assumption because in most conductance-based models of neurons and cardiac pacemaker cells, we can only give a single control input in the form of a current to one of the elements of the state vector, which corresponds to the voltage across the cell membrane.

\subsection{Desynchronizing Neurons} \label{desynch}
Parkinsonian and essential tremor affect millions of people worldwide, causing involuntary tremors in various parts of the body, and disrupting the activities of daily living. Pathological neural synchronization in the STN and the thalamus brain region is hypothesized to be one of the causes of motor symptoms of parkinsonian and essential tremor, respectively~\cite{STN_syncro,thalamus_syncro}. Deep brain stimulation (DBS), an FDA approved treatment, has proven to alleviate these symptoms~\cite{DBS_STN,DBS_thalamus} by stimulating the STN or the thalamus brain regions with a high frequency, (relatively) high energy pulsatile waveform, which has been hypothesized to desynchronize the synchronized neurons; see, e.g., \cite{wilson11,clustered}. This has motivated researchers to come up with efficient model dependent control techniques~\cite{tass2003,neural_control2,dan14} which not only desynchronize the neurons but also consume less energy, thus prolonging the battery life of the stimulator and preventing tissue damage or side effects caused by the pulsatile stimuli. 

\begin{figure}[!t]
\begin{center}
\includegraphics[width=0.9\textwidth]{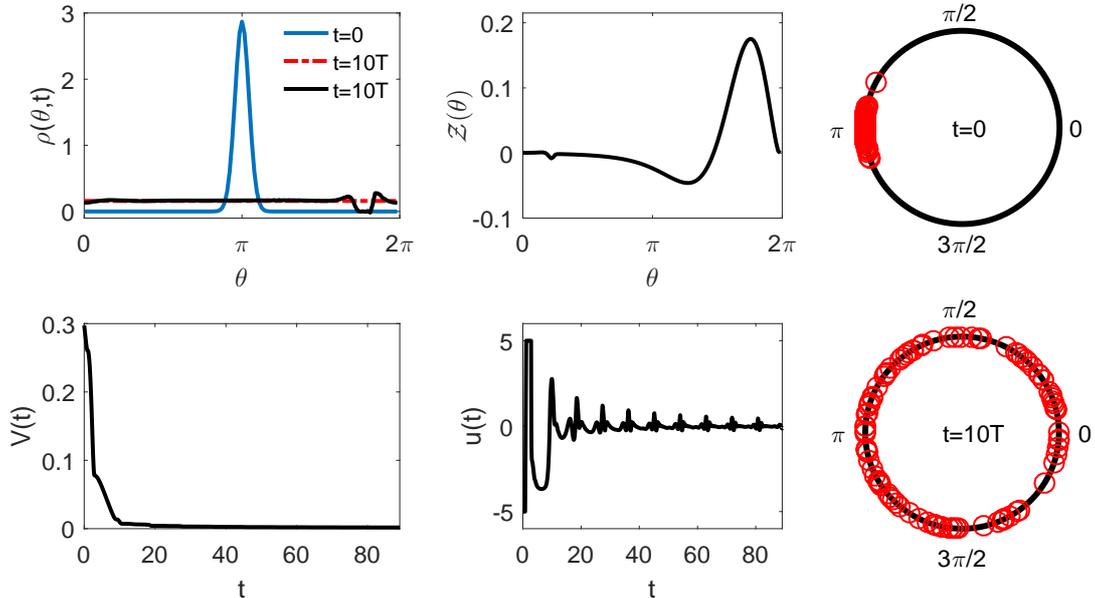}
\end{center}
\caption{Desynchronizing Control: In the top left panel, the solid (resp., red dashed) lines show the probability distribution $\rho(\theta,t)$ (resp., $\rho_f(\theta,t)$) at various times. The top middle panel shows the PRC, while the bottom left and middle panels show the Lyapunov function $V(t)$ (\ref{lf}), and the control input, respectively.  The top (resp., bottom) right panels show 100 phase oscillators at time $t=0~ms$ (resp., $t=10T~ms$). Here $T=8.91~ms$.}
\label{hh}
\end{figure}

Thus, inspired by this treatment of parkinsonian and essential tremor, we employ our algorithm to desynchronize an initially synchronized population of neurons. Here we use our algorithm to change the probability distribution of synchronized neurons with mean $\pi$ and $\kappa=52$, into a uniform distribution ($\kappa=0$). As a proof of concept, we use the two-dimensional reduced Hodgkin-Huxley model~\cite{Keener2009,huxley} for calculating the PRC. For details of this model, see \ref{a_hh_param}. Under zero control input, this model gives a stable periodic orbit with time period $T=8.91ms$. The top middle panel of Figure \ref{hh} shows the corresponding PRC. 
We take the control parameters $P=1000, \ u_{min}=-5, \ u_{max}=5$, and simulate until $t=10T$. From the top and bottom left panels of Figure \ref{hh}, we see that the control input is able to flatten out the bell shaped probability distribution, and thus decrease the Lyapunov function towards zero. For $t>10~ms$, we see that decay rate of Lyapunov function decreases, and thus the Lyapunov function asymptotically decreases towards zero. This can be explained from equations (\ref{dotofv} - \ref{pc}) where we see that control input (decay rate of Lyapunov function) depends on the (square of the) difference between current coefficients and desired coefficients. Thus as the coefficients get closer to their desired value, the magnitude of the control input decreases significantly, which decreases the rate of decay of the Lyapunov function. The top right panel of Figure \ref{hh} shows 100 phase oscillators synchronized with mean $\pi$, and $\kappa=52$ extracted through the Matlab circular statistical toolbox. We apply the control input from the middle bottom panel of Figure \ref{hh} to them in an open loop manner. The bottom right panel of the same figure shows the same oscillators at time $t=10T$. We see that the control input is able to desynchronize these oscillators almost perfectly. In transforming the probability distribution, the total control energy consumed ($\int_0^{10T} u^2 dt$) comes out to be $141.78$ units.

\subsection{Clustering Neurons for Maximizing Neural Plasticity} \label{cluster}
An adult human brain is composed of hundreds of billions of neurons, and each of these neurons is connected to other neurons. Neural plasticity is a significant factor in forming specific connections by wiring neurons that fire together \cite{firewire}. Spike time dependent plasticity (STDP) is one type of long-term plasticity, which wires neurons that fire together over a long period of time, thus helping in the regulation of neural synchrony. However, increased neural synchrony is a hallmark of several neurological disorders as discussed in the previous section, and STDP can resynchronize a desynchronized neural population over time in the presence of noise \cite{stdp2}. Thus, desynchronizing control, as considered in the previous section, may not be the best long-term solution. Recent results \cite{clustered} suggest another hypothesis that DBS works by forming neural clusters instead of complete desynchronization. Coordinated Reset, a method which promotes clustering, has shown to have long lasting effects even after the control stimulus is turned off \cite{tass2003,tass14}. This further motivates clustering as an alternative desynchronizing strategy for the treatment of parkinsonian and essential tremor. This would not only reduce neural synchrony but also promote clustering over long periods of time by re-wiring of neuron connections through STDP. We demonstrate this by first defining the phase difference distribution, and then the STDP curve.

\subsubsection{Phase Difference Distribution}
Given a phase distribution $\rho(\theta,t)$ governing the probability of a population of oscillators at phase $\theta$ and time $t$, a corresponding phase difference distribution $ \rho_{d}(\phi,t)$ governs the probability that the phase difference between any two set of oscillators in the population is $\phi$ at time $t$, where $\phi \in [0,2\pi)$. We only consider uncoupled oscillators which evolve independently from each other in this article. Thus the probability that the phase difference between any two oscillators is $\phi$ at time $t$ can be given by the integral of the products of the phase distribution and the phase distribution shifted by $\phi$ at times $t$ over the entire domain:
\begin{equation}
 \rho_{d}(\phi,t)=\int_0^{2\pi}\rho(\theta_s,t)\rho(\theta_s+\phi,t)d\theta_s.\label{pdiff2}
\end{equation}
The phase difference distribution satisfies
\begin{equation}
\int_0^{2\pi}\rho_{d}(\phi,t)d\phi=1.
\end{equation}
This can be shown from equation (\ref{pdiff2}):
\begin{eqnarray*}
\int_0^{2\pi}\rho_{d}(\phi,t)d\phi&=&\int_0^{2\pi} \left[\int_0^{2\pi}\rho(\theta_s,t)\rho(\theta_s+\phi,t)d\theta_s\right]d\phi\\
&=&\int_0^{2\pi} \left[ \int_0^{2\pi}\rho(\theta_s+\phi,t)d\phi \right]  \rho(\theta_s,t) d\theta_s\\
&=&\int_0^{2\pi} 1 \cdot \rho(\theta_s,t) d\theta_s\\
&=&1.
\end{eqnarray*}
Note that phase difference distribution for a time-dependent traveling wave $\rho_f(\theta,t)$ governed by equation (\ref{dfpe}), is stationary and does not depend on time. This can be proven by taking the time derivative of equation (\ref{pdiff2}):
\begin{eqnarray*}
\frac{d \rho_{d}}{dt}&=&\int_0^{2\pi}\left[\frac{\partial \rho_f(\theta_s,t)}{\partial t}\rho_f(\theta_s+\phi,t)+\rho_f(\theta_s,t)\frac{\partial \rho_f(\theta_s+\phi,t)}{\partial t}\right]d\theta_s\\
&=&-\omega\int_0^{2\pi}\left[\frac{\partial \rho_f(\theta_s,t)}{\partial \theta_s}\rho_f(\theta_s+\phi,t)+\rho_f(\theta_s,t)\frac{\partial \rho_f(\theta_s+\phi,t)}{\partial \theta_s}\right]d\theta_s\\
&=&-\omega\left.\rho_f(\theta_s,t)\rho_f(\theta_s+\phi,t)\right|_0^{2\pi}+\omega\int_0^{2\pi}\rho_f(\theta_s,t)\frac{\partial \rho_f(\theta_s+\phi,t)}{\partial \theta_s}d\theta_s\\
&&\qquad \qquad \qquad \qquad \qquad \qquad-\omega\int_0^{2\pi}\rho_f(\theta_s,t)\frac{\partial \rho_f(\theta_s+\phi,t)}{\partial \theta_s}d\theta_s\\
&=&0.
\end{eqnarray*}
Here, the first equality follows from the Leibniz rule from elementary calculus, and the third equality follows from the previous line by applying integration by parts and imposing periodic boundary conditions. Thus, this proves that the phase difference distribution for a time-dependent traveling wave  is independent of time. For such a traveling wave phase distribution, we write the phase difference distribution as being independent of time: 
\begin{equation}
 \rho_{d}(\phi)=\int_0^{2\pi}\rho_f(\theta_s,t)\rho_f(\theta_s+\phi,t)d\theta_s.\label{pdiff}
\end{equation}

The Fourier coefficients for the phase difference distribution can be calculated as follows:
\begin{eqnarray*}
\qquad \qquad  \rho_{d}(\phi)=\int_0^{2\pi} \left(\frac{1}{2\pi}+\right. \sum_{k=1}^{N-1} \left. \left[\widetilde{A_k}(t)\cos (k\theta_s) + \widetilde{B_k}(t) \sin (k \theta_s) \right] \right) \qquad \qquad \qquad \qquad  \\ 
\qquad \qquad   \times \left(\frac{1}{2\pi}+\sum_{l=1}^{N-1}  \left[\widetilde{A_l}(t)\cos (l(\theta_s+\phi)) + \widetilde{B_l}(t) \sin (l( \theta_s+\phi)) \right]\right) d\theta_s.
\end{eqnarray*}
By expanding $\cos (l(\theta_s+\phi))$ and $\sin (l(\theta_s+\phi))$, and making use of the orthogonality of $\cos k \theta_s$ and $\sin k \theta_s$, we obtain
\begin{equation}
\rho_{d}(\phi)=\frac{1}{2\pi}+\pi\sum_{k=1}^{N-1} \left(\widetilde{A_k}^2(t)+\widetilde{B_k}^2(t)\right)\cos (k\phi).\label{pdiff_fs}
\end{equation}
From this formulation of the phase difference distribution in terms of the Fourier coefficients of the desired phase distribution, one can easily verify that $\rho_{d}(\phi)$ is $2\pi$-periodic, $\int_0^{2\pi}\rho_{d}(\phi)d\phi=1$, and $\rho_{d}(\phi)$ is independent of time, which can be seen by taking the time derivative of equation (\ref{pdiff_fs}):
\begin{eqnarray*}
\dot \rho_{d}(\phi)&=\pi&\sum_{k=1}^{N-1} 2\left(\widetilde{A_k}(t) \dot {\widetilde{A_k}}(t)+\widetilde{B_k}(t) \dot {\widetilde{B_k}}(t)\right)\cos (k\phi)\\
&=&\pi\sum_{k=1}^{N-1} 2k\omega\left(-\widetilde{A_k}(t) \widetilde{B_k}(t)+\widetilde{A_k}(t) \widetilde{B_k}(t)\right)\cos (k\phi)\\&=&0
\end{eqnarray*}
Another property that the phase difference distribution has is that it always has a local maximum at zero phase difference. This can easily be verified from equation (\ref{pdiff_fs}), as $\frac{d\rho_{d}(0)}{d\phi}=0$ and $\frac{d^2\rho_{d}(0)}{d\phi^2}<0$.

\subsubsection{Spike Time Dependent Plasticity Stabilizes Clusters}
STDP is an asymmetric form of Hebbian learning \cite{hebb} that modifies synaptic connections between neurons when they fire repeatedly in a causal manner \cite{stdp0,stdp1,stdp4}. At the single synapse level, STDP potentiates (resp., depresses) the synaptic strength for repeated pre-synaptic action potentials arriving just before (resp., after) the post-synaptic action potential. At the population level, STDP strengthens the synaptic connections between neurons that fire action potentials synchronously and weakens those connections in the out of phase neurons \cite{stdp2}. Plasticity is known to be an important factor in the formation of neural pathways in initial brain development, as well as later in learning and memory storage. Since we consider uncoupled oscillating neurons in this article, we reformulate STDP in terms of the phase difference $\phi$ between two neurons instead of their spike time difference; the distribution of interspike intervals is same as the phase difference distribution for uncoupled oscillating neurons. If the phase difference $\phi \in [0,\pi)$, the STDP would increase the synaptic weight, and if the phase difference $\phi \in [\pi,2\pi)$, STDP would depress the synaptic weight. We call this increase or decrease of synaptic weights as a function of phase difference the {\it STDP curve} given as
\begin{eqnarray}
S(\phi)=\left\{\begin{array}{c} pe^{-\frac{\phi}{\tau_p}}, \qquad  \phi \in [0,\pi)\\ -de^{\frac{\phi-2\pi}{\tau_d}}, \qquad \phi \in [\pi,2\pi)\end{array} \right. .\label{stdp_curve}
\end{eqnarray}
\begin{figure}[!t]
\begin{center}
\includegraphics[width=0.8\textwidth]{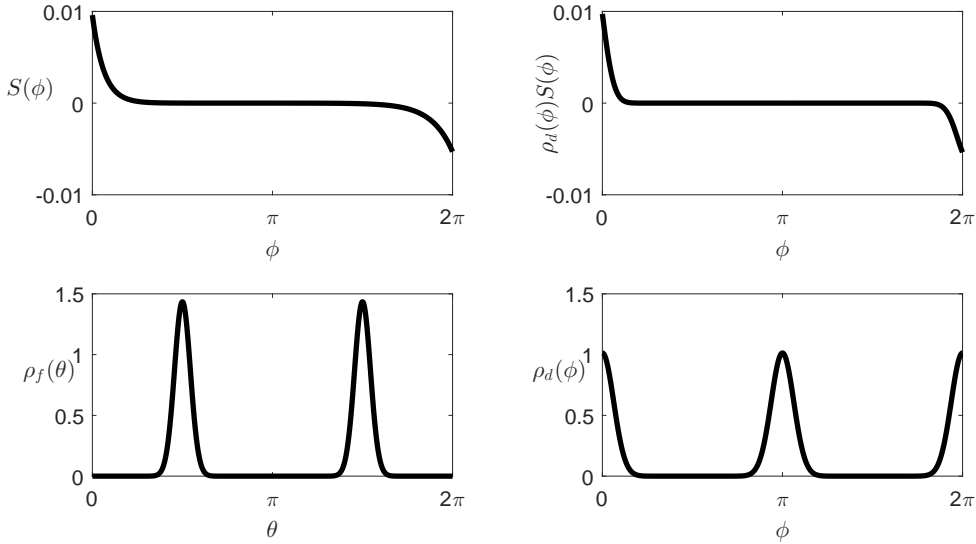}
\end{center}
\caption{The top left panel shows the spike time dependent plasticity curve $S(\phi)$. The bottom left (resp., right) panel shows the desired phase (resp., phase difference) distribution. The top right panel shows the change in synaptic weight between two neurons as a function of their phase difference.}
\label{phase_diff}
\end{figure}
We take the parameters $p=0.0096, \ d=0.0053$ from \cite{stdp1}, while $\tau_p=0.2, \ \tau_d=0.365$ are taken so that the integral of the resulting STDP curve ($\int_0^{2\pi}S(\phi)d\phi$) is zero \cite{stdp3}. The top left panel of Figure \ref{phase_diff} shows the STDP curve $S(\phi)$ with the above parameters.

\begin{figure}[!t]
\begin{center}
\includegraphics[width=0.9\textwidth]{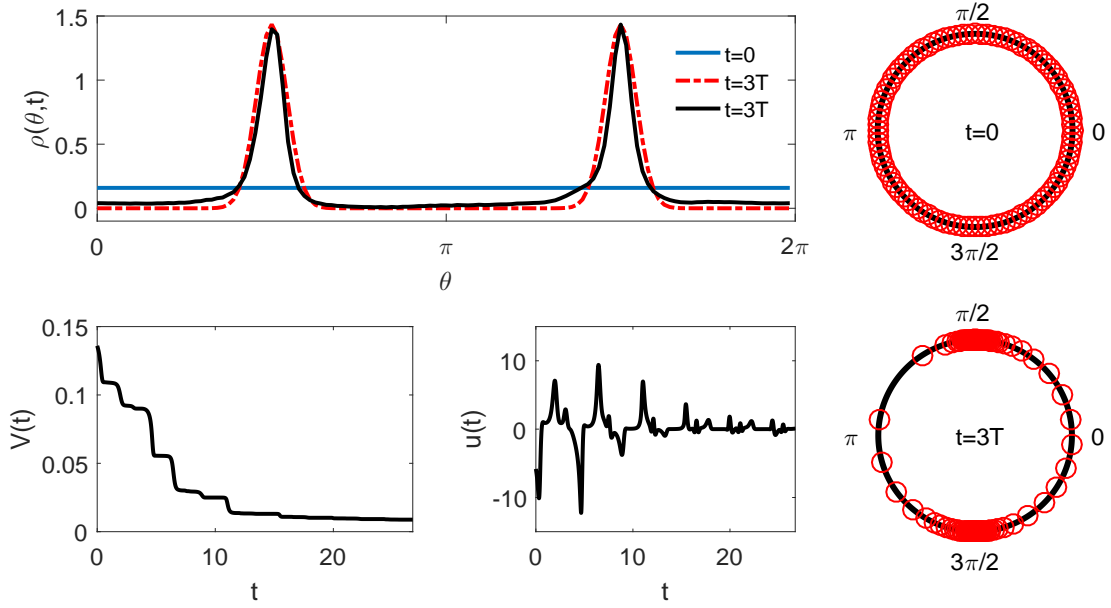}
\end{center}
\caption{Clustering Control: In the top left panel, the solid (resp., red dashed) lines show the probability distribution $\rho(\theta,t)$ (resp., $\rho_f(\theta,t)$) at various times. The bottom left and middle panels show the Lyapunov function $V(t)$ (\ref{lf}), and the control input, respectively.  The top (resp., bottom) right panels show 100 desynchronized (resp., clustered) phase oscillators at time $t=0~ms$ (resp., $t=3T~ms$). Here $T=8.91~ms$.}
\label{hh_cluster}
\end{figure}

Let us suppose that we start with a desynchronized population. The average rate of synaptic connection change between any two neurons in the population is given by \cite{stdp2}
\begin{equation}
\Delta c=\int_0^{2\pi}\rho_{d}(\phi)S(\phi)d\phi.\label{synapse_change}
\end{equation}
A uniform phase distribution (desynchronized population) would result in a uniform phase difference distribution, which would lead to a zero average synaptic change. On the other hand, if we promote neural clustering, STDP would potentiate intra-cluster synaptic connections and depress inter-cluster connections. This would thus potentially help in long-term stabilizability of clusters in the presence of noise. We demonstrate this by taking two clusters and calculating the average synaptic change (\ref{synapse_change}) intra- and inter-cluster. Thus we take the desired phase distribution as a bi-modal distribution, which can be realized as a sum of two uni-modal von Mises distributions
\begin{eqnarray}
\rho_f(\theta,t)=\frac{e^{\kappa \cos (\theta+\pi/2)}+e^{\kappa \cos (\theta+3\pi/2)}}{4\pi \mathcal{I}_0(\kappa)},\label{bimo}
\end{eqnarray}
where $\kappa=52$. From this we calculate the phase difference distribution from equation (\ref{pdiff}) or (\ref{pdiff_fs}), which can then be used to calculate the average synaptic change from equation (\ref{synapse_change}). The bottom left, right, and top right panels of Figure \ref{phase_diff} show the desired phase distribution, phase difference distribution, and the product of the phase difference distribution with the STDP curve respectively. The average synaptic change for intra- and inter-cluster is calculated as
\begin{eqnarray}
\Delta c_{intra-cluster}&=&\int_0^{\frac{\pi}{2}}\rho_{d}(\phi)S(\phi)d\phi + \int_{\frac{3\pi}{2}}^{2\pi}\rho_{d}(\phi)S(\phi)d\phi = 3.62\times 10^{-4},\\
\Delta c_{inter-cluster}&=& \int_{\frac{\pi}{2}}^{\frac{3\pi}{2}}\rho_{d}(\phi)S(\phi)d\phi = -3.96\times 10^{-7}.
\end{eqnarray}
Thus STDP would strengthen synapse in each cluster and weaken them between the two clusters, thereby potentially maintaining clusters over a long period of time. This motivates us to transform an initially desynchronized phase distribution ($\kappa=0$) into a bi-modal phase distribution (\ref{bimo}). As a proof of concept, here we again use the two-dimensional reduced Hodgkin-Huxley model for calculating the PRC. We take the control parameters $P=1200, \ u_{min}=-15, \ u_{max}=15$, and simulate until $t=15T$. The results are shown in Figure \ref{hh_cluster}.
From the top and bottom left panels of Figure \ref{hh_cluster}, we see that the control input is able to transform an initial uniform distribution into a bi-modal distribution, and thus the Lyapunov function decreases towards zero. As the Fourier coefficients of the current and desired distribution get closer, the control input decreases in magnitude, which decreases the rate of decay of the Lyapunov function. The top right panel of Figure \ref{hh_cluster} shows 100 desynchronized phase oscillators ($\Theta_i(0)=2\pi(i-1)/100$) to which the control input from the middle bottom panel of Figure \ref{hh_cluster} is applied in an open loop manner. The bottom right panel of the same figure shows the oscillators at time $t=3T$. We see that the control input is able to separate the desynchronized oscillators into two distinct clusters corresponding to the bi-modal phase distribution. In transforming the probability distribution, the total control energy consumed ($\int_0^{3T} u^2 dt$) comes out to be $152.59$ units.

\subsection{Eliminating Cardiac Alternans}\label{yni_param}
The collection of cells in the Sinoatrial node called cardiac pacemaker cells elicit periodic electrical pulses which polarize a collection of excitable and contractile cells called myocytes. In the process of depolarizing, myocytes contract and propagate action potentials to the neighboring cells.  This well-coordinated process of excitation / depolarization and contraction enables the heart to pump blood throughout the body. Under normal conditions, with constant pacing by the cardiac pacemaker cells, the action potential duration (APD), that is the time for which an action potential lasts in a myocyte cell, also remains constant. However, under some conditions, this 1:1 rhythm between pacing and the APD can become unstable, bifurcating into a 2:2 rhythm of alternating long and short APD, known as alternans~\cite{alternans}.  Alternans is observed to be a possible first step leading to fibrillation~\cite{fibrillation}. Thus, a number of researchers have worked on suppressing alternans as a method of preventing fibrillation, thereby preventing the need for painful and damaging defibrillating shocks. Many of these methods~\cite{alter_control1,alter_control2,alter_control3,alter_control4} operate by exciting the myocardium tissue externally with periodic pulses, and changing the period according to the alternating rhythm. However, such a control requires excitation at several sites in the tissue~\cite{multiple_sites}. 

\begin{figure}[!t]
\begin{center}
\includegraphics[width=0.9\textwidth]{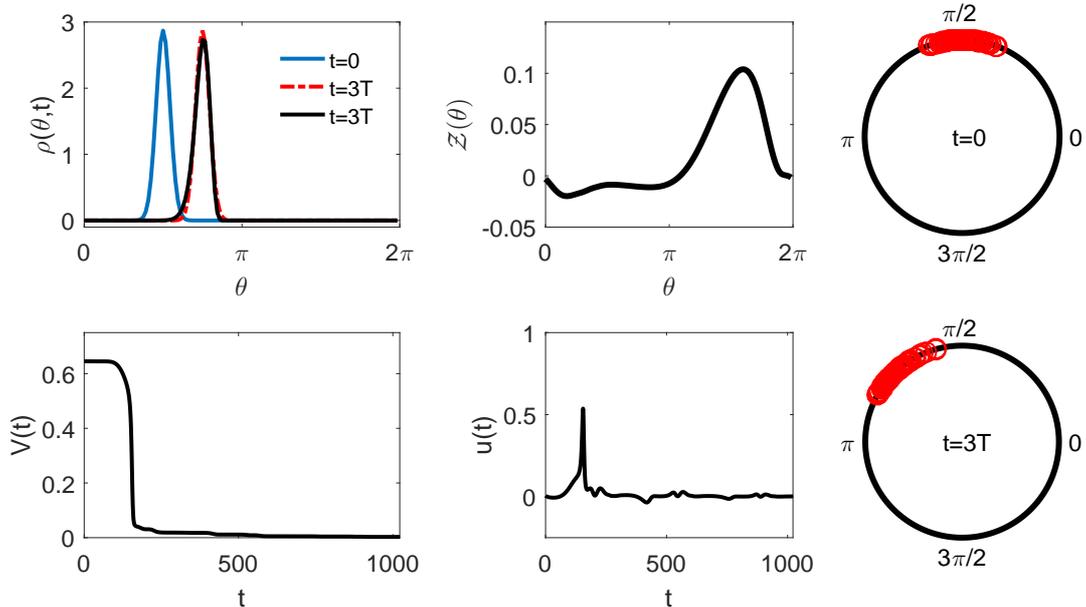}
\end{center}
\caption{Phase shifting cardiac pacemaker cells: In the top left panel, the solid (resp., red dashed) lines show the probability distribution $\rho(\theta,t)$ (resp., $\rho_f(\theta,t)$) at various times. The top middle panel shows the PRC, while the bottom left and middle panels show the Lyapunov function $V(t)$ (\ref{lf}), and the control input, respectively.  The top (resp., bottom) right panels show 100 phase oscillators at time $t=0~ms$ (resp., $t=3T~ms$). Here $T=340.8~ms$. Note that in the absence of control input, the oscillators would have a mean of $\pi/2$ at $t=3T$.}
\label{alternan}
\end{figure}

In \cite{monga}, we developed a novel strategy to suppress alternans by changing the phase of the pacemaker cells. The control strategy was based on a single oscillator model to change the phase of a single cell. However for an effective cardiac alternans elimination, we need to consider the entire population of cardiac pacemaker cells which oscillate in synchrony. So, here we aim to phase shift the population of cardiac pacemaker cells using the control algorithm we developed in Section \ref{control_design}. Such a control strategy could eliminate the need to excite the tissue at multiple sites. The amount of phase change required to eliminate alternans depends on the discrete APD dynamics \cite{monga}. Here we advance the phase by $\pi/4$ as an example. For the PRC calculation, we consider phase reduction of the 7-dimensional YNI model of SA node cells in rabbit heart proposed in \cite{yni}. The model is of Hodgkin-Huxley type with 6 gating variables and a transmembrane voltage variable on which the control input acts. For details of the model, see \ref{a_yni_param}. With this model we get a stable periodic orbit with time period $T=340.8~ms$. We start with a synchronous population distribution with mean $\pi/2$ and $\kappa=52$. In order to phase shift this distribution by $\pi/4$, we take the target population distribution $\rho_f(\theta,t)$ with same $\kappa$ value but an initial mean of $3\pi/4$. Thus our control algorithm will push the distribution $\rho(\theta,t)$ forward until it matches with the desired distribution $\rho_f(\theta,t)$. We take the control parameters $P=5, \ u_{min}=-1, \ u_{max}=1$ and apply control input until $t=3T$. From the top and bottom right panels of Figure \ref{alternan}, we see that the control input is able to phase shift the probability distribution, and thus decreases the Lyapunov function towards zero. In doing so, it changes the shape of the distribution slightly. The top right panel of Figure \ref{alternan} shows 100 phase oscillators synchronized with mean $\pi/2$, and $\kappa=52$ extracted through the Matlab circular statistical toolbox. We apply the control input from the middle bottom panel of the figure to those in an open loop manner. The bottom right panel of the same figure shows the oscillators at time $t=3T$. We see that the control input is able to phase shift these oscillators by $\pi/4$. The slight change in shape of the phase distribution is reflected in the final position of phase oscillators, where handful of the oscillators get spread relative to the main cluster. In shifting the phase of the probability distribution, the total control energy consumed ($\int_0^{3T} u^2 dt$) comes out to be $3.21$ units.
 
Here we mention another application for which shifting the phase of an oscillator population is desired: phase shifting circadian oscillators for the treatment of jet lag. Neurons in the suprachiasmatic nucleus (SCN) of the brain are responsible for maintaining the circadian rhythm in mammals. This rhythm is synchronized with the external day and night cycle under normal conditions.  A disruption between these two rhythms can happen due to multiple reasons, such as travel across time zones, starting a night shift job, working in extreme environments (space, earth poles, underwater), etc. Such asynchrony can lead to several physiological disorders \cite{circ_health,circ_clinical}, thus motivating researchers to try to develop ways to remove it. In \cite{monga}, we developed a strategy to eliminate this asynchrony by changing the phase of a single SCN neuron by using a light stimulus as the control input, since light is known to affect the circadian rhythm~\cite{light_circ1}.  This would change the phase of the circadian rhythm so that it gets aligned with the external cycle after the end of the controlled oscillation. However for a better alignment of the circadian rhythm with the external environment, we need to phase shift the entire population of SCN neurons which oscillate in synchrony, which can be achieved by our control algorithm. This is very similar to the previous application of phase shifting cardiac pacemaker cells.

\section{Addition of White Noise}\label{coup_noise}
So far we have demonstrated that our control is effective for a population of uncoupled, noise-free oscillators. However, real systems are subjected to noise; thus, in this section we modify our control to take this into account. 

Given $M$ noisy, uncoupled oscillators with dynamics given by
\begin{equation}
\frac{d \mbf{x}_j}{dt} = F(\mbf{x}_j) + \left[ u(t) + \sqrt{2D}\eta_j(t),\ 0,\ \ldots,\ 0\right]^T, \quad j=1,\ldots,M.
\end{equation}
Here each oscillator receives a common input $u(t)$ modified by a different realization of Gaussian white noise $\sqrt{2D}\eta_j(t)$ with zero mean, variance $2D$, and with $\langle \eta_i(t) \eta_j(t') \rangle = \delta_{ij}\delta(t-t')$. Letting $\theta_j$ be the phase of the $j^{th}$ oscillator, to leading order in the noise strength Ito's formula gives \cite{gardiner2004}
\begin{equation}
\dot \theta_j = \omega + \mathcal{Z}(\theta)\left[ u(t) + \sqrt{2D}\eta_j(t)\right], \quad j=1,\ldots,M.\label{phase_noise}
\end{equation}
Assuming $M$ is large and noise perturbations are small, the population dynamics can be represented in terms of its phase distribution $\rho(\theta,t)$ with stochastic averaging \cite{st_avg,dan_pde}:
\begin{equation}
\frac{\partial \rho(\theta,t)}{\partial t}=-\frac{\partial }{\partial \theta}\left[\left(\omega+\mathcal{Z}(\theta) u(t)\right)\rho(\theta,t)\right] + \frac{B^2}{2}\frac{\partial^2 \rho(\theta,t)}{\partial \theta^2},\label{phase_dis_noise}
\end{equation}
where 
\begin{equation}
B^2 = \frac{2D}{2\pi}\int_0^{2\pi} \mathcal{Z}^2(\theta) d\theta.\nonumber
\end{equation}
As before, the desired final probability distribution $\rho_f(\theta,t)$ is taken to be a traveling wave which evolves according to equation (\ref{dfpe}). To devise our control laws, we use the approximation of a finite Fourier series to write the phase distributions (equations (\ref{rho}), and (\ref{rhof})). The Fourier coefficients of the desired phase distribution evolve as before (equations (\ref{afk}), and (\ref{bfk})), whereas the Fourier coefficients of phase distribution evolve as
\begin{eqnarray}
 \dot A_k(t) &=& -k\omega B_k  - \mathcal{I}_{kA}(t)u(t)  -\frac{B^2}{2}k^2A_k(t),\\
 \dot B_k(t) &=& k\omega A_k  + \mathcal{I}_{kB}(t)u(t)-\frac{B^2}{2}k^2B_k(t).
\end{eqnarray}

\subsection{Control Design}
Here as well we take the Lyapunov function as the sum of squared differences of the Fourier coefficients of the current and the desired distributions (equation (\ref{lf})). Its derivative in time evolves as
\begin{equation}
\dot V(t)=I(t)u(t) + G(t),
\end{equation}
where
\begin{eqnarray*}
G(t) =-\frac{B^2}{2} \sum_{k=1}^{N-1}k^2 \left[A_k(t) \left(A_k(t)-\widetilde{A_k}(t)\right)+ B_k(t)\left(B_k(t)-\widetilde{B_k}(t)\right)\right],
\end{eqnarray*}
and $I(t)$ is given by equation (\ref{integrall}). Then by taking the control input
\begin{eqnarray}
u(t)= -PI(t)-\frac{G(t)}{I(t)},  \label{pc_noise}
\end{eqnarray}
where $P$ is a positive scalar, we get the time derivative of the Lyapunov function to be a negative scalar. Thus, according to the Lyapunov theorem, the control law (\ref{pc_noise}) will decrease the Lyapunov function until the current probability distribution becomes equal to the desired distribution. Here we do not consider the degenerate case where $I(t)\equiv 0$ when $\rho(\theta,t)\ne\rho_f(\theta,t)$ (see Section \ref{degeneracy} for cases when $I(t)\equiv 0$ when $\rho(\theta,t)\ne \rho_f(\theta,t)$). 

\subsection{Simulation Results}
\begin{figure}[!t]
\begin{center}
\includegraphics[width=0.9\textwidth]{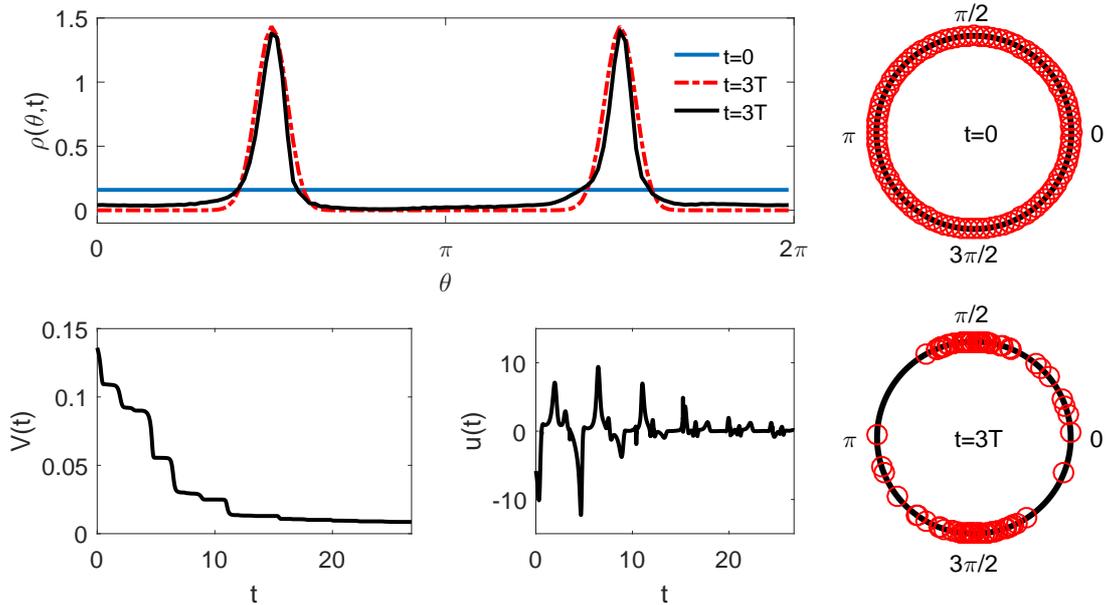}
\end{center}
\caption{Clustering Control in presence of noise: In the top left panel, the solid (resp., red dashed) lines show the probability distribution $\rho(\theta,t)$ (resp., $\rho_f(\theta,t)$) at various times. The bottom middle and left panels show the control input (\ref{pc_noise}), and the Lyapunov function $V(t)$ (\ref{lf}), respectively.  The top (resp., bottom) right panels show 100 desynchronized (resp., clustered) phase oscillators at time $t=0~ms$ (resp., $t=3T$ ms). Here $T=8.91~ms$.}
\label{hh_cluster_noise}
\end{figure}

To demonstrate our control in the presence of noise, we use (\ref{pc_noise}) to transform an initial uniform phase distribution into a desired bi-modal distribution (\ref{bimo}). We take the noise strength $\sqrt{2D}=0.03$ in equations (\ref{phase_dis_noise}) and (\ref{phase_noise}). To simulate the noisy phase oscillators, we write the equation (\ref{phase_noise}) as
\begin{equation*}
d \theta_j = \omega dt + \mathcal{Z}(\theta)\left[ u(t) dt+ \sqrt{2D}dW_j(t)\right], \quad j=1,\ldots,M,
\end{equation*}
where $dW_j(t) =\eta_j(t)dt$ and $W_j(t)$ is the standard Weiner process. We use the second order Runge-Kutta algorithm developed in \cite{honeycutt} to simulate the above equation, and use \mbox{randn} with a predefined seed in Matlab for generating the standard Weiner process. In order to be consistent, we evaluate the phase distribution and the control input using a second order Runge-Kutta method as well. As a proof of concept, here we again use the two-dimensional reduced Hodgkin-Huxley model for calculating the PRC. We take the control parameter $P=1200$, and simulate until $t=3T$. The results are shown in Figure \ref{hh_cluster_noise}.
From the top panel, we see that the control input is able to transform an initial uniform distribution into a bi-modal distribution, and thus the Lyapunov function decreases towards zero. The top right panel of Figure \ref{hh_cluster_noise} shows 100 desynchronized phase oscillators to which the control input from the bottom middle panel of Figure \ref{hh_cluster_noise} is applied in an open loop manner. As seen from the bottom right panel of Figure \ref{hh_cluster_noise}, the control input is able to separate the desynchronized oscillators into two distinct clusters corresponding to the bi-modal phase distribution. In transforming the probability distribution, the total control energy consumed ($\int_0^{3T} u(t)^2 dt$) comes out to be $153.30$ units.  The control input $u(t)$ used for this energy consumption calculation is taken from equation (\ref{pc_noise}), and thus is same for all stochastic realizations with the same noise intensity. The energy consumption is $0.46\%$ more than the similar control without noise. This is expected as the addition of white noise introduces a diffusion term in the phase distribution PDE, and thus causes the phase distribution to decay down towards a uniform distribution with time. Therefore, the control has to expend additional effort in transforming the phase distribution into a bi-modal distribution. We note that non-zero control will be necessary to maintain the bi-modal distribution for all time.

\section{Optimal Control of Phase Distributions}\label{optimal_control}
In this section we formulate an optimal control algorithm to transform the phase distribution $\rho(\theta,t)$ into the desired distribution $\rho_f(\theta,t)$. We do this by controlling the Fourier coefficients of the phase distribution. We start with the coefficients $A_k(0)$ and $B_k(0)$ of $\rho(\theta,t)$ at time $t=0$, and want them to match the coefficients $\widetilde{A_k}(\tau)$ and $\widetilde{B_k}(\tau)$ of $\rho_f(\theta,t)$ at time $t=\tau$. Thus we pose the optimal control problem as the following Two Point Boundary Value Problem (BVP). We take the cost function $R$ as 
\begin{equation}
R=\int_0^{\tau} \left\{u^2 +\sum_{k=1}^{N-1}\left[\lambda_{kA}\left(\dot A_k + k\omega B_k + \mathcal{I}_{kA} u \right) +\lambda_{kB}\left(\dot B_k  - k\omega A_k - \mathcal{I}_{kB} u \right) \right]\right\}dt.
\label{aop}
\end{equation}
The first term in the cost function ensures that the control law uses a minimum energy input. The second term ensures that the phase distribution evolves according to equation (\ref{fpe}), with $\lambda_{kA}$ and  $\lambda_{kB}$ being  the Lagrange multipliers. The Euler-Lagrange equations are obtained from
 \begin{equation}
\frac{\partial P}{\partial q}=\frac{d}{d t}\left( \frac{\partial P}{\partial \dot q}\right), \qquad q=\lambda_{kA}, \ \lambda_{kB},\ A_k, \ B_k, \ u,
\end{equation}
where $P$ is the integrand in the cost function $R$. This gives the Euler-Lagrange equations for $k=1,\ldots,N-1$:
\begin{eqnarray}
\dot A_k&=&-k\omega B_k - \mathcal{I}_{kA}u,\\
\dot B_k&=&k\omega A_k +\mathcal{I}_{kB}u,\\
\dot \lambda_{kA} &=& -k\omega \lambda_{kB} +\mathcal{H}_{kA}u,\\
\dot \lambda_{kB} &=& k\omega \lambda_{kA} +\mathcal{H}_{kB}u,
\end{eqnarray}
where
 \begin{eqnarray}
u&=&\frac{1}{2} \sum_{k=1}^{N-1} \left[ \lambda_{kB}\mathcal{I}_{kB} - \lambda_{kA}\mathcal{I}_{kA} \right],\label{control_opt}\\
\mathcal{H}_{kA}&=&\frac{1}{\pi} \int_0^{2\pi} \mathcal{Z}(\theta) \Lambda(\theta,t) \cos(k\theta)  d \theta,\\
\mathcal{H}_{kB}&=&\frac{1}{\pi} \int_0^{2\pi} \mathcal{Z}(\theta) \Lambda(\theta,t) \sin(k\theta)  d \theta,\\
 \Lambda(\theta,t) &=& \sum_{l=1}^{N-1} l\left[ \lambda_{lA} \sin(l \theta) -\lambda_{lB} \cos(l \theta) \right ].
\end{eqnarray}
We solve the Euler-Lagrange equations as a two point BVP with the boundary conditions:
\begin{eqnarray} \begin{array}{l} 
A_k(0)=\frac{1}{\pi}\int_0^{2\pi}\rho(\theta,0)\cos (k\theta) d\theta, \ \ B_k(0)=\frac{1}{\pi}\int_0^{2\pi}\rho(\theta,0)\sin (k\theta) d\theta, \vspace{2mm} \\  A_k(\tau)=\frac{1}{\pi}\int_0^{2\pi}\rho_f(\theta,\tau)\cos (k\theta) d\theta, \ B_k(\tau)=\frac{1}{\pi}\int_0^{2\pi}\rho_f(\theta,\tau)\sin (k\theta) d\theta. \end{array}
\end{eqnarray}
Since $A_k(0)$, and $B_k(0)$ are fixed by the problem, the BVP can be solved by finding appropriate values of $\lambda_{kA}(0)$ and $\lambda_{kB}(0)$. We formulate a modified Newton iteration method to solve this high dimensional ($2N-2$) BVP. For details of the method, see \ref{bvp_mni}. 

\begin{figure}[!t]
\begin{center}
\includegraphics[width=0.9\textwidth]{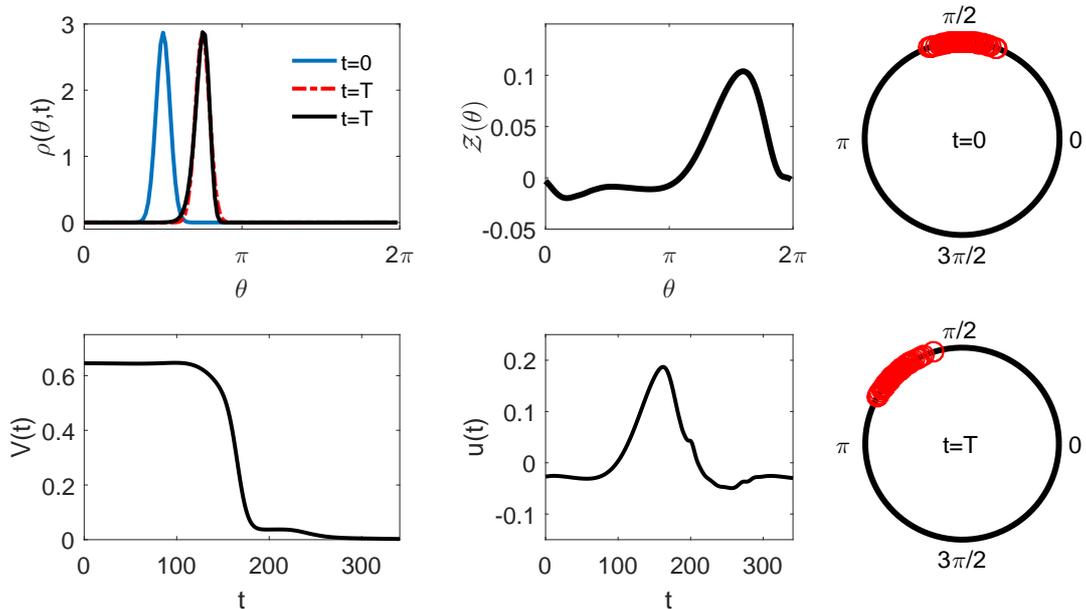}
\end{center}
\caption{Phase shifting cardiac pacemaker cells through optimal control: In the top left panel, the solid (resp., red dashed) lines show the probability distribution $\rho(\theta,t)$ (resp., $\rho_f(\theta,t)$) at various times. The top middle panel shows the PRC, while the bottom left and middle panels show the Lyapunov function $V(t)$ (\ref{lf}), and the control input, respectively.  The top (resp., bottom) right panels show 100 phase oscillators at time $t=0~ms$ (resp., $t=T~ms$). Here $T=340.8~ms$. Note that in the absence of control input, the oscillators would have a mean of $\pi/2$ at $t=T$.}
\label{optimal}
\end{figure}

We demonstrate the control by considering the application of phase shifting a distribution as we did in Section \ref{yni_param}. Here as well we consider the YNI model of SA node cells in rabbit heart. We start with a synchronous population distribution with mean $\pi/2$ and $\kappa=52$. We use our optimal control algorithm to phase shift this distribution by $\pi/4$ in time $\tau=T$. So, we take the target distribution $\rho_f(\theta,t)$ with same $\kappa$ value but an initial mean of $3\pi/4$. We also compute the Lyapunov function $V(t)$ for comparison with our results from Section \ref{yni_param}, even though our optimal control algorithm is not based on this Lyapunov function. Results are shown in Figure \ref{optimal}. From the top and bottom left panels of Figure \ref{optimal}, we see that the control input is able to phase shift the phase distribution, and thus decreases the Lyapunov function towards zero (non-monotonically in this case). The top right panel of Figure \ref{optimal} shows 100 phase oscillators synchronized with mean $\pi/2$, and $\kappa=52$ extracted through the Matlab circular statistical toolbox. We apply the control input from the middle bottom panel of the figure to them in an open loop manner. The bottom right panel of the same figure shows the oscillators at time $t=T$. We see that the control input is able to phase shift these oscillators by $\pi/4$. In shifting the phase of the probability distribution, the total control energy consumed ($\int_0^{T} u^2 dt$) comes out to be $1.56$ units, which is less than half of the energy required for the same control objective using our Lyapunov-based control algorithm in Section \ref{yni_param}. We note that this energy comparison is fair as in both cases the control input decreases the Lyapunov function by the same amount (by 99.6\%). Thus our optimal control is able to achieve the control objective while simultaneously minimizing the amount of total energy required.

\section{Conclusion}\label{con}
In this article we developed a framework to control a population of uncoupled oscillators by controlling their phase distributions. By writing the phase distribution as a finite Fourier series, we were able to decompose the PDE governing the evolution of the phase distribution into a set of ODEs governing the evolution of the corresponding Fourier coefficients. We formulated our control algorithms in Fourier space as well, driving the Fourier coefficients of the current phase distribution to the Fourier coefficients of the desired distribution with a single control input. For our first Lyapunov-based control algorithm, we constructed a degenerate set of phase distributions and the phase response curves in terms of their Fourier coefficients. We extended this algorithm to take into account the effect of white noise on the dynamics of the oscillator population. Finally, we formulated an optimal control algorithm which uses a minimum energy input to achieve the desired phase distribution. Our control algorithms are quite flexible; for the systems considered in this paper, they have the potential to drive a system of uncoupled oscillators from any initial phase distribution to any traveling-wave final phase distribution, as long as the combination of those distributions is non-degenerate.

We demonstrated the versatility of our control algorithms by using them for three distinct applications. First, motivated by the hypothesis of neural synchronization in the STN and the thalamus brain region as one of the causes of motor symptoms of parkinsonian and essential tremor, respectively, we applied our control algorithm to drive an initial synchronous phase distribution to a uniform distribution. For the second application, we defined the phase difference distribution in terms of the phase distribution, and proved some of its fundamental properties. This formulation of the phase difference distribution was essential in demonstrating the importance of a clustered neural population for enhancing spike time dependent plasticity, and thus re-wiring of neural connections for better stability of the partially synchronous clustered state. Motivated by the elimination of cardiac alternans, we applied our algorithm to control a population of synchronized cardiac pacemaker cells by advancing their phase distribution by a specified phase. For all these applications, we demonstrated the effectiveness of our control by applying the respective control inputs to a population  of 100 uncoupled noise-free phase oscillators.

We conclude with remarks about the experimental implementation of these algorithms. Since they require knowledge of the current Fourier coefficients of the phase distribution, one would need to measure neuronal/cardiac pacemaker cell activity in order to back out the phase distribution and hence the Fourier coefficients in real time. This measurement would require good spatial and temporal resolution, so for both neuroscience and cardiovascular experiments we suggest that the use of Micro-Electrode arrays (MEA) would be a good fit. Note that for in vivo experiments, fMRI and EEG are unlikely to be the right tools since fMRI has poor temporal resolution, while EEG is susceptible to noise and poorly measures neural activity beneath the cortex. An experimental setup in general will include effects due to coupling, which are absent in our control algorithm. Our algorithm would still work on such systems as long as the coupling is weak. If synchrony is stable with coupling, then it would be harder for our control algorithm to desynchronize a synchronized population in the presence of coupling. The addition of noise might make this even harder if STDP is present, as STDP promotes synchrony in the presence of noise. However, in the absence of STDP, noise would make it easier for our control algorithm to desynchronize a synchronized oscillator population.

\appendix
\section{Models}\label{model_param}
In this appendix, we give details of the mathematical models used in this article.

\subsection{Reduced Hodgkin-Huxley model}\label{a_hh_param}
Here we list the reduced Hodgkin-Huxley model \cite{canard,Keener2009,huxley} used in Section \ref{desynch}:
\begin{eqnarray*}
\dot v&=& \left(I - g_{Na}(m_\infty)^3(0.8-n)(v-v_{Na}) - g_Kn^4(v-v_K)-g_L(v-v_L)\right)/c+u(t),\\ 
\dot n &=& a_n(1-n)-b_n n,
\end{eqnarray*}
where $v$ is the trans-membrane voltage, and $n$ is the gating variable. $I$ is the baseline current, for which we use the units $\mu A/cm^2$, and $u(t)$ represents the applied control current.
\begin{eqnarray*}
a_n &=&0.01(v+55)/(1-\exp(-(v+55)/10)),\\ 
b_n &=& 0.125\exp(-(v+65)/80),\\
a_m &=& 0.1(v+40)/(1-\exp(-(v+40)/10)),\\
b_m &=& 4\exp(-(v+65)/18),\\
m_\infty &=& a_m / (a_m + b_m),\\
c &=& 1,\   g_L = 0.3,\      g_{Na} = 120,\    v_{Na} = 50,\\g_K &=36& ,\    v_K = -77,\     v_L = -54.4,\ I =20 .
\end{eqnarray*}
Here, $\theta=0$ corresponds to the initial condition $v=42.8828,\ n=0.4920$.

\subsection{YNI model}\label{a_yni_param}
Here we list model parameters of the YNI model \cite{yni} introduced in Section \ref{yni_param}. It is given as
\begin{eqnarray*}
\dot v&=&-\frac{I_{Na}+I_k+I_l+I_s+I_h}{C}+u(t),\\
\dot d&=&\alpha_d(1-d)-\beta_d d,\\
\dot f&=&\alpha_f(1-f)-\beta_f f,\\
\dot m&=&\alpha_m(1-m)-\beta_m m,\\
\dot h&=&\alpha_h(1-h)-\beta_h h,\\
\dot q&=&\alpha_q(1-q)-\beta_q q,\\
\dot p&=&\alpha_p(1-p)-\beta_p p,
\end{eqnarray*}
where $v$ represents the transmembrane voltage, and $d,f,m,h,q,p$ are the gating variables, $u(t)$ represents the applied current as the control input, with parameters
\begin{eqnarray*}
\alpha_d&=&\frac{0.01045(v+35)}{(1-\exp(-(v+35)/2.5))+\frac{0.03125v}{(1-\exp(-v/4.8))}},\\
\beta_d&=&0.00421(v-5)/(-1+\exp((v-5)/2.5)),\\
\alpha_f&=&0.000355(v+20)/(-1+\exp((v+20)/5.633)),\\
\beta_f&=&0.000944(v+60)/(1+\exp(-(v+29.5)/4.16)),\\
\alpha_m&=&(v+37)/(1-\exp(-(v+37)/10)),\\
\beta_m&=&40\exp(-0.056(v+62)),\\
\alpha_h&=&0.001209(\exp(-(v+20)/6.534)),\\
\beta_h&=&1/(1+\exp(-(v+30)/10)),\\
\alpha_q&=&0.0000495+\frac{0.00034(v+100)}{(-1+\exp((v+100)/4.4))},\\
\beta_q&=&0.0000845+0.0005(v+40)/(1-\exp(-(v+40)/6)),\\
\alpha_p&=&0.0006+0.009/(1+\exp(-(v+3.8)/9.71)),\\
\beta_p&=&0.000225(v+40)/(-1+\exp((v+40)/13.3)),\\
i_s&=&12.5(\exp((v-30)/15)-1),\\
I_s&=&(0.95d+0.05)(0.95f+0.05)i_s,\\
I_{Na}&=&0.5m^3h(v-30),\\
I_h&=&0.4q(v+25),\\
I_k&=&0.7p(\exp(0.0277(v+90))-1)/\exp(0.0277(v+40)),\\
I_l&=&0.8(-\exp(-(v+60)/20)+1),\\
C&=&1.
\end{eqnarray*}
Here, $\theta=0$ corresponds to the initial condition $v=-19.2803, \ d= 0.6817, \    f=0.0236, \    m=0.8540, \    h=0.0013, \    q=0.0038, \   p= 0.6592$.

\section{Two point BVP with modified Newton Iteration}\label{bvp_mni}
We solve the Euler-Lagrange equations as a two point boundary value problem using a modified Newton iteration method, which we briefly summarize. Consider a general two point boundary value problem
\begin{equation}
\dot y = f(t,y), \qquad y \in \mathbb{R}^n, \qquad 0\le t \le \tau,\label{bvp}
\end{equation}
with the linear boundary condition
\begin{equation*}
B_0y(0)+B_\tau y(\tau)=a, \qquad B_0,\ B_\tau \in \mathbb{R}^{n\times n}.
\end{equation*}
To solve such a boundary value problem, we integrate equation (\ref{bvp}) with the initial guess $c=y(0)$, and calculate the function $g(c)$:
\begin{equation*}
g(c)=B_0c+B_\tau y(\tau)-a,
\end{equation*}
where $y(\tau)$ is the solution at time $\tau$ with the initial condition $c$. If we had chosen the correct initial condition $c$, $g(c)$ would be 0. Based on the current guess $c^\nu$, and the $g(c^\nu)$ value, we choose the next initial condition by the modified Newton Iteration as an element-wise update 
\begin{equation}
c^{\nu+1}_i=c^\nu_i -\left(\left.\frac{\partial g_i}{\partial c_i}\right|_{c^\nu}\right)^{-1}g_i(c^\nu), \ \ \text{for} \ i=1, \ldots , n\label{NI}
\end{equation}
where $g_i$, and $c_i^\nu$ represent the $i^{th}$ element of vectors $g$, and $c^\nu$ respectively. We compute the derivative $J_{ii}=\left.\frac{\partial g_i}{\partial c_i}\right|_{c^\nu}$ numerically as
\begin{equation*}
J_{ii}=\frac{g^+_i - g^-_i}{2\epsilon},
\end{equation*}
where 
\begin{eqnarray*}
g^+_i&=&g_i\left(c^\nu+e_i \epsilon\right),\\
g^-_i&=&g_i\left(c^\nu-e_i \epsilon\right),
\end{eqnarray*}
$\epsilon$ is a small number, and $e_i$ is a column vector with 1 in the $i^{\rm th}$ position and 0 elsewhere.
\subsection{Solving Euler-Lagrange equations}
For the Euler-Lagrange equations devised in Section \ref{optimal_control}, $A_k(0)$, and $B_k(0)$ are fixed by the initial distribution, so the only way to control the distribution is by choosing appropriate values of $\lambda_{kA}(0)$ and $\lambda_{kB}(0)$. Thus our BVP can be reduced to $2N-2$ dimensions even though the Euler-Lagrange equations are $4N-4$ dimensional. The $i^{th}$ element of the vector $c$ is taken as
\begin{eqnarray*}
c_i=\left\{ \begin{array}{l} \lambda_{kA}(0), \quad \text{for} \ i=k=1,\ldots,N-1 \\  \lambda_{kB}(0),  \quad \text{for} \ i=k+N-1=N,\ldots,2N-2. \end{array}\right.
\end{eqnarray*}
The $i^{th}$ element of the vector $g(c)$ for $i=k=1,\ldots,N-1$ is taken as
\begin{eqnarray*}
g_i(c)= A_k(0) + A_k(\tau) -\frac{1}{\pi}\int_0^{2\pi}\left(\rho(\theta,0)+\rho_f(\theta,\tau) \right)\cos (k\theta) d\theta, 
\end{eqnarray*}
and, for $i=k+N-1=N,\ldots,2N-2$,
\begin{eqnarray*}
g_i(c)=  B_k(0) + B_k(\tau) -\frac{1}{\pi}\int_0^{2\pi}\left(\rho(\theta,0) + \rho_f(\theta,\tau) \right) \sin (k\theta) d\theta. 
\end{eqnarray*}
The derivative $J_{ii}$ is given as
\begin{eqnarray*}
\frac{\partial g_i}{\partial c_i}=\left\{\begin{array}{l} \frac{\partial A_k(\tau)}{\partial \lambda_{kA}(0)}, \quad \text{for} \ i=k=1,\ldots,N-1 \\  \frac{\partial B_k(\tau)}{\partial \lambda_{kB}(0)}, \quad \text{for}  \ i=k+N-1=N,\ldots,2N-2.\end{array}\right.
\end{eqnarray*}
This information is used in equation (\ref{NI}) to iteratively find the appropriate value of the vector $c$. 




\section*{ACKNOWLEDGMENT}
This work was supported by National Science Foundation Grant No. NSF-1635542. We thank Prof. Gary Froyland for helpful discussions during the initial stage of this project. We thank Timothy Matchen for pointing out the degeneracy issue, and Raphael Egan for helpful discussions on the pseudospectral method.

\bibliographystyle{siamplain}
\bibliography{ms}

\end{document}